\documentclass[conference]{IEEEtran}
\IEEEoverridecommandlockouts

\usepackage{hyperref}                        
\hypersetup{
    colorlinks=true,
    citecolor=black,
    linkcolor=black,
    linkbordercolor={1 1 1},
    citebordercolor={1 1 1},    
}
\usepackage{amssymb}                         
\usepackage{amsmath}                         
\usepackage{booktabs}                        
\usepackage{graphicx}
\usepackage[sort]{cite}

\usepackage{epstopdf}
\usepackage{float}
\usepackage{soul}
\usepackage{color}
\usepackage[inline, shortlabels]{enumitem} 

\usepackage{balance}

\newcommand{\Equation}[1]{\eqref{#1}}

\newcommand{\Figure}[1]{Figure~\ref{#1}}
\newcommand{\Figures}[2]{Figures~\eqref{#1} and~\eqref{#2}}
\newcommand{\Table}[1]{Table~\ref{#1}}

\newcommand{\Section}[1]{Section~\ref{#1}}
\newcommand{\Sections}[2]{Sections~\ref{#1}~and~\ref{#2}}

\usepackage[numbers]{natbib}
\usepackage{notoccite}
\usepackage{multirow}

\DeclareMathAlphabet{\pazocal}{OMS}{zplm}{m}{n}

\newcommand{\Na}{\pazocal{N}}

\newcommand{\Ra}{\pazocal{R}}
\newcommand{\Ia}{\pazocal{I}}

\usepackage{subcaption}

\usepackage{mathtools}

\usepackage{lipsum}
\usepackage{acronym}
\acrodef{SV}{speaker verification}
\acrodef{TCN}{temporal convolutional network}
\acrodef{DNN}{deep neural network}
\acrodef{RIR}{room impulse response}
\acrodef{DET}{detection error tradeoff}
\acrodef{EER}{equal error rate}

\def\BibTeX{{\rm B\kern-.05em{\sc i\kern-.025em b}\kern-.08em
    T\kern-.1667em\lower.7ex\hbox{E}\kern-.125emX}}

\begin{document}

\title{Speaker Verification in Multi-Speaker Environments Using Temporal Feature Fusion}

\author{
\IEEEauthorblockN{Ahmad Aloradi, Wolfgang Mack, Mohamed Elminshawi, Emanu{\"e}l A.~P. Habets}
\IEEEauthorblockA{International Audio Laboratories Erlangen\textsuperscript{*}\thanks{*A joint institution of the Friedrich-Alexander-University Erlangen-N{\"u}rnberg (FAU) and Fraunhofer Institute for Integrated Circuits (IIS).}, Am Wolfsmantel 33, 91058 Erlangen, Germany \\
\{ahmad.aloradi,wolgang.mack,mohamed.elminshawi,emanuel.habets\}@audiolabs-erlangen.de}
}

\maketitle

\begin{abstract}
Verifying the identity of a speaker is crucial in modern human-machine interfaces, e.g., to ensure privacy protection or to enable biometric authentication. Classical \ac{SV} approaches estimate a fixed-dimensional embedding from a speech utterance that encodes the speaker’s voice characteristics. 
A speaker is verified if his/her voice embedding is sufficiently similar to the embedding of the claimed speaker.
However, such approaches assume that only a single speaker exists in the input. The presence of concurrent speakers is likely to have detrimental effects on the performance. To address \ac{SV} in a multi-speaker environment, we propose an end-to-end deep learning-based \ac{SV} system that detects whether the target speaker exists within an input or not.
First, an embedding is estimated from a reference utterance to represent the target's characteristics. Second, frame-level features are estimated from the input mixture.
The reference embedding is then fused frame-wise with the mixture's features to allow distinguishing the target from other speakers on a frame basis.
Finally, the fused features are used to predict whether the target speaker is active in the speech segment or not. Experimental evaluation shows that the proposed method outperforms the x-vector in multi-speaker conditions. 
\end{abstract}

\acresetall

\section{Introduction}
\label{sec:Introduction}

\Ac{SV} is a binary classification task that aims at verifying the claimed identity of a speaker. The pipeline of classical \ac{SV} systems consists of a front-end that projects speech onto speaker-discriminative embeddings followed by a back-end to compute the verification scores. 
Training speaker-discriminative embeddings is commonly done by optimizing classification or metric learning losses.
In the metric learning approach, embeddings are trained to capture the similarities between utterances of the same or different speakers by optimizing a metric learning objective, e.g., the triplet loss \cite{8352546} or the contrastive loss \cite{6857341}.
In the classification approach, the system is first trained to capture the speakers' characteristics by classifying speakers in the training set \cite{snyder2017deep, 8461375}.
At inference time, generalization to unseen speakers is achieved by adopting bottleneck features as embeddings.

The most prominent baseline of the classification-based systems is the x-vector system,  which can be currently considered state-of-the-art in \ac{SV} \cite{Lin2020,  mak_chien_2020}.
The rising popularity of the x-vector inspired various works to improve upon its architecture and training procedure.
For example, the frame-level feature extraction was advanced by including elements of the ResNet architecture \cite{7780459}.
The statistics pooling,  which aggregates the frame-level mean and standard deviation, was also refined by incorporating different attention mechanisms \cite{Desplanques2020,  zhu2018self,  Villalba2019}.
Moreover, the vanilla softmax loss was replaced by more sophisticated classification losses that yielded improved performance \cite{Zhou2019}.

Most research in \ac{SV} assumes that there is only
one speaker per input segment, which is reflected by the majority of the standard evaluation datasets \cite{8683760}. 
However, real-world recordings might violate the premise of a single speaker.
The abundance of single-speaker datasets encouraged embracing this assumption and developing systems accordingly.
Nowadays, state-of-the-art \ac{SV} generally functions by capturing the speaker's characteristics in fixed-dimensional embeddings.
If there are multiple speakers to represent, the effectiveness of the learned embeddings is reduced \cite{8683760}. 
In such scenarios, it is common to use speaker diarization (i.e., to identify who spoke when \cite{6135543}) to improve the verification performance \cite{8683760,  Villalba2019, Lee2019}.

However, diarization is better suited in scenarios for which speakers are only partially overlapping. 
The emphasis in the present study is on significant overlaps between speakers, which may generally resemble mixtures in speech separation. 
In such conditions, a speaker is verified if his or her speech is present in a segment containing speech from multiple speakers \cite{Martin}. 

We hypothesize that answering the question "does the target speaker exist in the mixture or not?" requires modifying the traditional \ac{SV} paradigm. Specifically, utterances do not need to be mapped onto embeddings that are trained to discriminate between speakers. Instead, we propose to focus on identifying commonalities between speech from the target speaker and the mixture. Information about the target speaker can be obtained from the speaker's enrollment data.

To address speaker verification in a multi-speaker environment, we propose a binary classification \ac{DNN} that detects the presence of the target speaker based on a frame-level fusion of the target and the mixture features. The proposed system takes a pair of inputs: the mixture and a reference speech uttered by the target speaker. First, the reference and the mixture signals are processed using two different networks to extract their frame-level features. For the reference speech, these features are aggregated temporally to form an embedding that encodes the characteristics of the target.
This embedding is then fused with the time-variant features of the mixture. Subsequently, the fused features are used to verify whether the target speaker is active in the mixture or not. Contrary to the previous works, the proposed approach specifically addresses the concurrent speakers problem in a direct manner without entailing speaker diarization, speaker separation, or an enhancement front-end.

The remainder of the paper is structured as follows: In \Section{sec:Related_work}, we review a state-of-the-art \ac{SV} approach which is later used as a baseline. In \Section{sec:Proposed_Method}, we introduce the proposed approach. In \Section{sec:Datasets_and_Setup}, we discuss the data generation and experimental setup. In \Section{sec:eval},  we evaluate the proposed method and discuss the results. Finally, in \Section{sec:conclusion}, we conclude the paper.

\section{Classical Speaker Verification}
\label{sec:Related_work}
\begin{figure}
\centering
\includegraphics[width=\columnwidth]{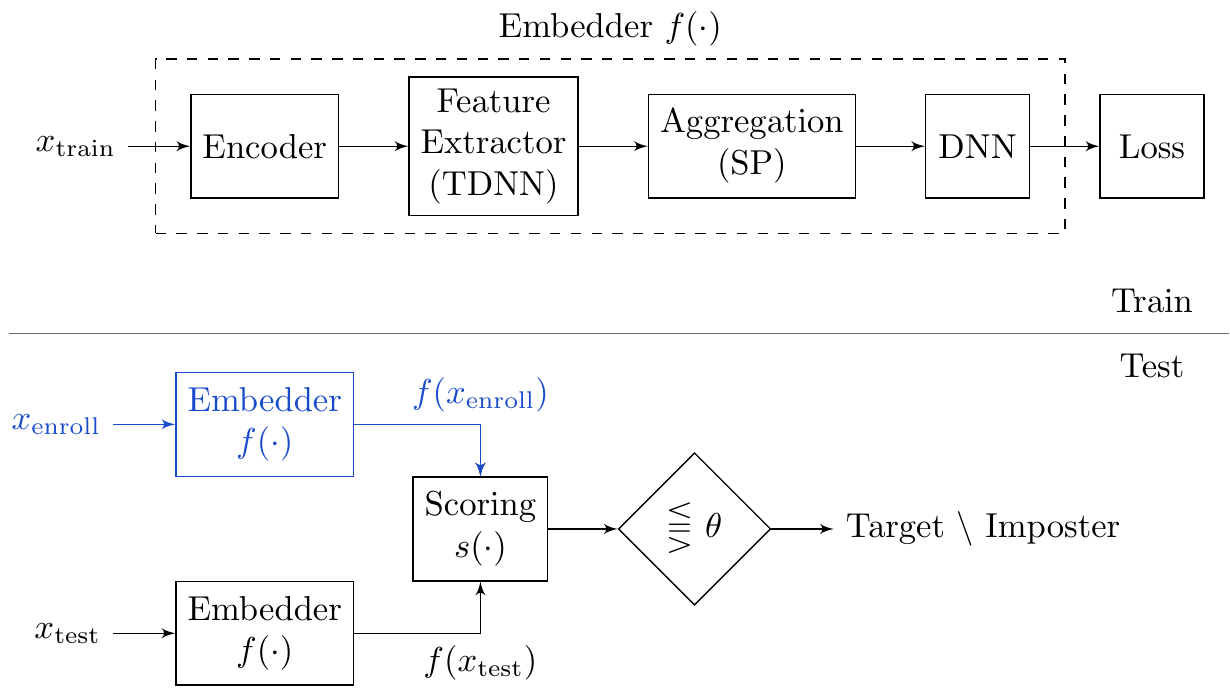}
\caption{\Ac{SV} approach based on the x-vector baseline. In the training phase, $f(\cdot)$ is trained by optimizing a loss to discriminate between speakers. After training, speakers are registered given their enrollment data $x_{\text{enroll}}$. At test time,  $f(x_{\text{test}})$ is computed and scored via $s(\cdot)$  against $f(x_{\text{enroll}})$ of the claimed target to determine if the speaker is the same or not.}
\label{fig:class1}
\end{figure}

Existing systems address the SV problem in two stages.  
First, embeddings are trained to discriminate between the speech signals based on the speaker's identity. 
After the training is finished, a scoring backend is used to determine whether the enrollment-time embedding of the target speaker matches the test-time embedding or not.
This high-level picture is depicted in \Figure{fig:class1}.

The SV problem formulation in the classical settings can be stated as follows.
Let  $f: \mathbb{R}^T \mapsto \mathbb{R}^D$ be the embedding mapping, and $s: \mathbb{R}^{D} \times \mathbb{R}^{D} \mapsto  \mathbb{R}$ be a similarity (scoring) function,  where $T$ is the number of time-domain samples and $D$ is the embeddings dimensionality. The embeddings $f$ are assumed to be learned by a neural network, and $s$ can be a predefined or trainable function.
The goal is to find $f$ and $s$ such that
\begin{equation}
\label{eqn:condition}
\begin{array}{r@{}l}
s(f(x^i_{\text{enroll}}),  f(x^j_{\text{test}})) \geqslant \theta &{} \;\; \text { if } j=i  ,  \\
s(f(x^i_{\text{enroll}}),  f(x^j_{\text{test}})) <  \theta &{} \;\; \text{  }  \forall j \neq i, \\

\end{array}
\end{equation}
where $x_{\text{enroll}} \in \mathbb{R}^{T_\text{enroll}}$ is the pre-recorded enrollment utterance used to model the target speaker, $x_{\text{test}} \in \mathbb{R}^{T_\text{test}}$ is the test-time utterance that needs to be verified, $\theta \in \mathbb{R}$ is the decision threshold,  and $i, j$ are speakers' labels. 

An issue with the aforementioned approach emerges when $x_{\text{test}}$ is a mixture containing speech from multiple concurrent speakers. That is, 
\begin{equation}
\displaystyle
x_{\text{test}} = x^i + \sum_{j \in \mathcal{Q}} x^j + v,
\label{eqn:x_test}
\end{equation}
where $\mathcal{Q}$ is a set of speaker labels for which $\mathcal{Q} \cup i = \emptyset$, and $v \in \mathbb{R}^{T_\text{test}}$ is an additive non-speech interference. In the following, we assume that $x^i$ and at least one $x^j$ with $j \neq i$ are concurrently active with comparable energy levels.
In this case, reliable detection of the $i$-th speaker from a set of $|\mathcal{Q}|+1$ speakers from $f(x_{\text{test}})$ is difficult since $f(x_{\text{test}})$ provides an embedding for the entire mixture and not only for the $i$-th speaker.
We hypothesize that in order to improve the target speaker detection, information about the target speaker (i.e., $x^i_{\text{enroll}}$) should be involved before producing the utterance-level representation, i.e., prior to the temporal aggregation step.
This hypothesis is the foundation of the proposed approach presented in \Section{sec:Proposed_Method}.




\section{Proposed Method}
\label{sec:Proposed_Method}

\begin{figure*}[t]
\centering
\includegraphics[width=0.95\textwidth]{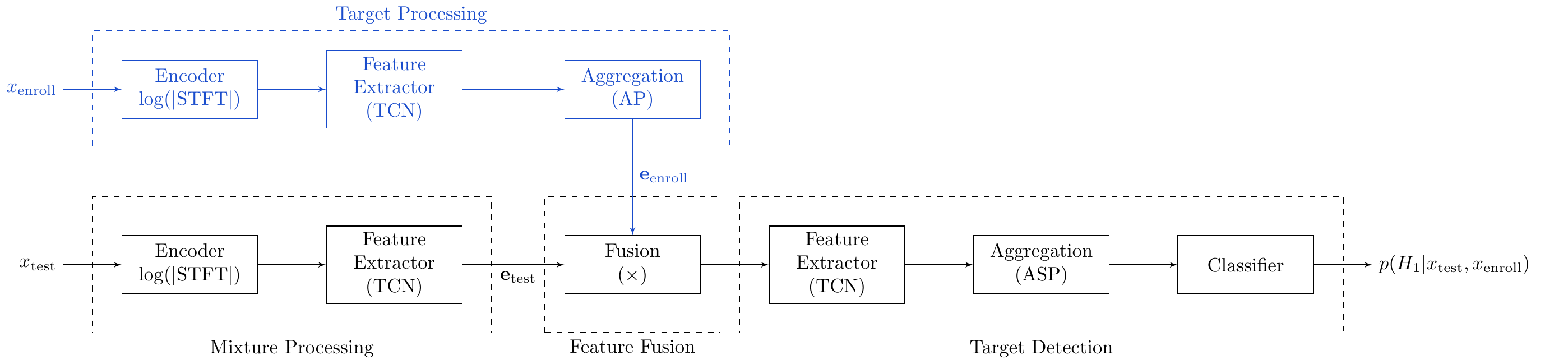}
\caption{Proposed temporal feature fusion SV system. For clarity, $\mathbf{e}_\text{enroll} \in \mathbb{R}^{N}$ and $\mathbf{e}_\text{test} \in \mathbb{R}^{N} \times \mathbb{R}^{L}$, where $N$ is the number of short-time Fourier transform (STFT) bases and $L$ is the number of time-frames of $x_\text{test}$ in the STFT domain.}

\label{fig:proposed_system}
\end{figure*}

\subsection{Problem Formulation}
To generalize to multi-speaker environments, we formulate the \ac{SV} problem as a detection problem. Namely, let us assume that the \ac{SV} system takes a pair of inputs $x^i_{\text{enroll}}$ and $x_{\text{test}}$ given by \Equation{eqn:x_test}. 
Based on this, we define the following hypotheses: 
\begin{equation} 
\label{eqn:proposed_cond}
\begin{split}
\text{Hypothesis } H_1 \text{:} \hspace{5pt} & i\mbox{-th speaker is present}~(x^i \neq 0)\\
\text{Hypothesis } H_0 \text{:} \hspace{5pt} & i\mbox{-th speaker is absent}~(x^i = 0).
\end{split}
\end{equation}
The goal of the proposed system is to find a model $g$ that estimates the target presence probability 
\begin{equation}
   g(x_{\text{test}} , x^i_{\text{enroll}}) = p(H_1|x_{\text{test}} , x^i_{\text{enroll}}).
\end{equation}
In this formulation, the network's output directly corresponds to the detection score. Therefore, the scoring backend $s$ is not required.

\subsection{System Overview}
The idea behind the proposed approach is to include target-specific information in the system before temporally aggregating the frame-level features. 
Features of the target speaker and the mixture are extracted from $x^i_{\text{enroll}}$ and $x_{\text{test}}$, respectively, fused to produce target-enhanced features, and then passed to the aggregation and other processing units, as shown in \Figure{fig:proposed_system}.
We believe it is essential to perform the feature fusion before temporal aggregation to emphasize the frame-level features with dominant target activity. 
Pooling can then be performed on the target-enhanced features to aggregate the statistics over multiple frames.
The described logic applies to the features of the mixture, but not to the features of the enrollment data that contains information about the target speaker only.
Hence, it is more intuitive to perform feature fusion using time-varying mixture features and time-invariant target features.
This design choice is also found in other systems that are used to address related problems \cite{wang2018voicefilter, 8736286}.

In our work,  we assume that the reference information of the target contains mostly clean speech. 
This assumption is valid at test time as the target speech can be obtained from the speaker's enrollment data.
\Sections{sec:model}{sec:train_and_test} describe the employed model and the proposed training and testing protocols, respectively.

\subsection{Model}
\label{sec:model}
The proposed model can be conceptualized in three steps: 
\begin{enumerate*}
\item[1)] target and mixture feature extraction, \item[2)] temporal feature fusion, and \item[3)] target speaker detection.
\end{enumerate*}
\Figure{fig:proposed_system} depicts the model structure and 
\Table{tab:arch} details the architecture. The frame-level features from the reference and the mixture signals are extracted using \ac{TCN} models, which have been shown to be successful feature extractors \cite{8707065,  9414765}. 
The learned features are extracted from the log spectrograms of the inputs.
In this step, only the frame-level features of the target are aggregated into an utterance-level representation $\mathbf{e}_{\text{enroll}}$ using average pooling (AP). After that, the extracted features are fused by element-wise multiplication. 
The fused features are processed frame-wise via a third \ac{TCN} , aggregated using attentive statistics pooling (ASP) \cite{zhu2018self}, and finally passed into a classifier to predict if the target is detected or not.

\begin{table}[t]
\centering
\caption{Architectural details of the proposed model with approx. 0.959 M parameters. The \acp{TCN} hyperparameters naming convention follow \cite{8707065}. The LinearBlock is a linear layer followed by ReLU and batch normalization.}
\label{tab:arch}
\begin{tabular}{@{}lll@{}}
\toprule
Module & Hyperparameters & Value(s)\\ \midrule
\acp{TCN}  & $\{N, B, H, P, X, R  \}$ & $\{257, 32, 64, 3 , 6, 3 \}$ \\
\hline
ASP & Attention channels & 128 \\
\hline
\multirow{4}{*}{Classifier} 
& Linear & (514, 257) \\
& LinearBlock & (257, 257) \\
& LinearBlock & (257, 257) \\
& Linear + Sigmoid & (257, 1) \\
\bottomrule
\end{tabular}
\end{table}

\subsection{Training and Testing}
\label{sec:train_and_test}
We used the binary cross-entropy (BCE) loss to train the proposed model. Assuming that the pair label $y$ equals $1$ when $H_1$ is true and $0$ otherwise, the training loss is defined as
\begin{multline}
  L_{\textrm{BCE}} = (y-1) \log(1-p(y|x^i_{\text{enroll}},  x_{\text{test}})) \\ - y \log(1-p(y|x^i_{\text{enroll}},  x_{\text{test}})).
\label{eqn:training_objective}  
\end{multline}
During training, $x_{\text{enroll}}$ and $x_{\text{test}}$ are sampled from the training data, and not from the enrollment and test data, respectively.

The inference process is naturally very similar to the training, except that no loss needs to be computed. That is, detecting whether or not the target speaker exists within the mixture is directly determined by the network's output.

\section{Experimental Setup and Data Generation}
\label{sec:Datasets_and_Setup}

The data in the experiments is based on VoxCeleb1 \cite{voxceleb1} and VoxCeleb2 \cite{voxceleb2}. The total number of speakers in the combined development sets is 7205 and the speech has a 16~kHz sampling rate. The development set is split into $90\%$ training and $10\%$ validation. Speech segments in either set have a fixed length of 3~seconds.
Each sample is augmented by noise and reverberation to produce three additional samples. The first augmented sample is a reverberant version of the speech signal. The second augmented sample is generated by adding noise from MUSAN \cite{musan}. The last augmented sample is generated by adding another speaker randomly selected from the current batch. Interference in the augmented samples was artificially reverberated with a probability of $0.2$. All the \aclp{RIR} we used are described in \cite{augmentation}. The signal-to-interference ratio (SIR) and signal-to-noise ratio (SNR) are uniformly sampled from the interval of $[0, 15]$.

For evaluation, we use three different test sets. The first is the Voxceleb1 test set (cleaned), which we refer to as the raw set $\Ra$. The second, referred to as $\Na$, is derived from $\Ra$ by corrupting it with additive non-speech from FSDnoisy18k test set \cite{8683158}. The third, referred to as $\Ia$, is derived from $\Ra$ by corrupting it with an interfering speaker sampled randomly from the validation set. The SNR and the SIR ranges are both uniformly sampled from the interval of $[0, 5]$. The number of interfering speakers is set to $1$.

Our experiments were based on the PyTorch framework \cite{NIPS2019_9015}. For the proposed model, the data was encoded using the log spectrogram with 512 discrete Fourier-transform points and a window length of 32~ms with 50\% overlap. We used the Adam optimizer \cite{kingma2014adam} with a learning rate $1e^{-4}$. The learning rate was reduced on a plateau by a factor $0.5$ every epoch the error rate did not improve on the validation set. The number of epochs was 20 and the batch size is 42.

The x-vector system in \cite{speechbrain} is used as a baseline. Furthermore, for a fair comparison, we trained another x-vector using the same training set and augmentations used to train the proposed model.
In the trained x-vector, we maintained almost identical hyperparameters as in the pre-trained system. However, due to the hardware limitations, we reduced the batch size in the trained model from 256 to 150. The SV trials of the x-vector were scored using cosine similarity.

\section{Performance Evaluation}
\label{sec:eval}
We report the results using the  \ac{EER}. Both the proposed model and the baseline were tested on the three sets $\Ra$, $\Na$, and $\Ia$. \Table{tab:results} outlines the results on the three test sets.
\begin{table}[!t]
\centering
\caption{\ac{EER}[\%] under various settings. Sets $\Ra$ is the Voxceleb1-test set (Cleaned) while $\Na$ and $\Ia$ are contaminated versions as described in \Section{sec:Datasets_and_Setup}.}
\label{tab:results}
\resizebox{0.8\columnwidth}{!}{
\begin{tabular}{@{}lccc@{}}
\toprule
\multirow{2}{*}{System}                                           & \multicolumn{3}{c}{Test Set} \\ \cmidrule(l){2-4} 
         & $\Ra$     & $\Na$     & $\Ia$     \\ \midrule
X-vector & 9.05 & 11.77 & 19.48  \\
X-vector (pre-trained) & 8.87    & 11.41   & 20.60   \\

Proposed & \textbf{7.81} & \textbf{8.70}  & \textbf{13.58} \\ \bottomrule
\end{tabular}
}
\end{table}

\begin{figure}[!tb]
    \centering
    
    \begin{subfigure}[t]{0.48 \columnwidth}
    \centering
    \caption{Official VoxCeleb1 splits $\Ra$}
    \includegraphics[width=\columnwidth]{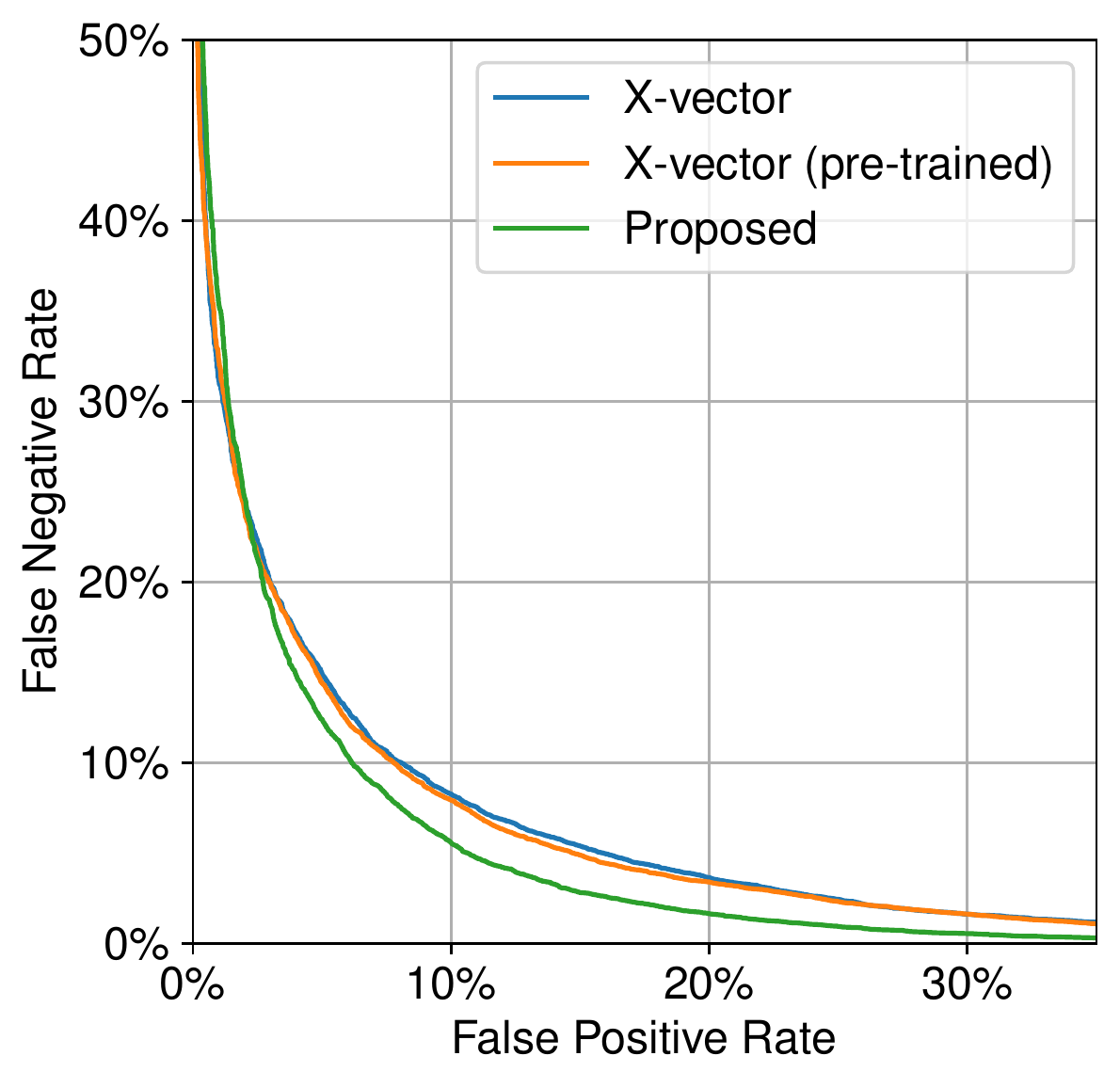}
    \label{fig:DET_C}
    \end{subfigure}
    ~ 
    \begin{subfigure}[t]{0.48 \columnwidth}
    \centering
    \caption{Multi-speaker scenario $\Ia$}
    \includegraphics[width=\columnwidth]{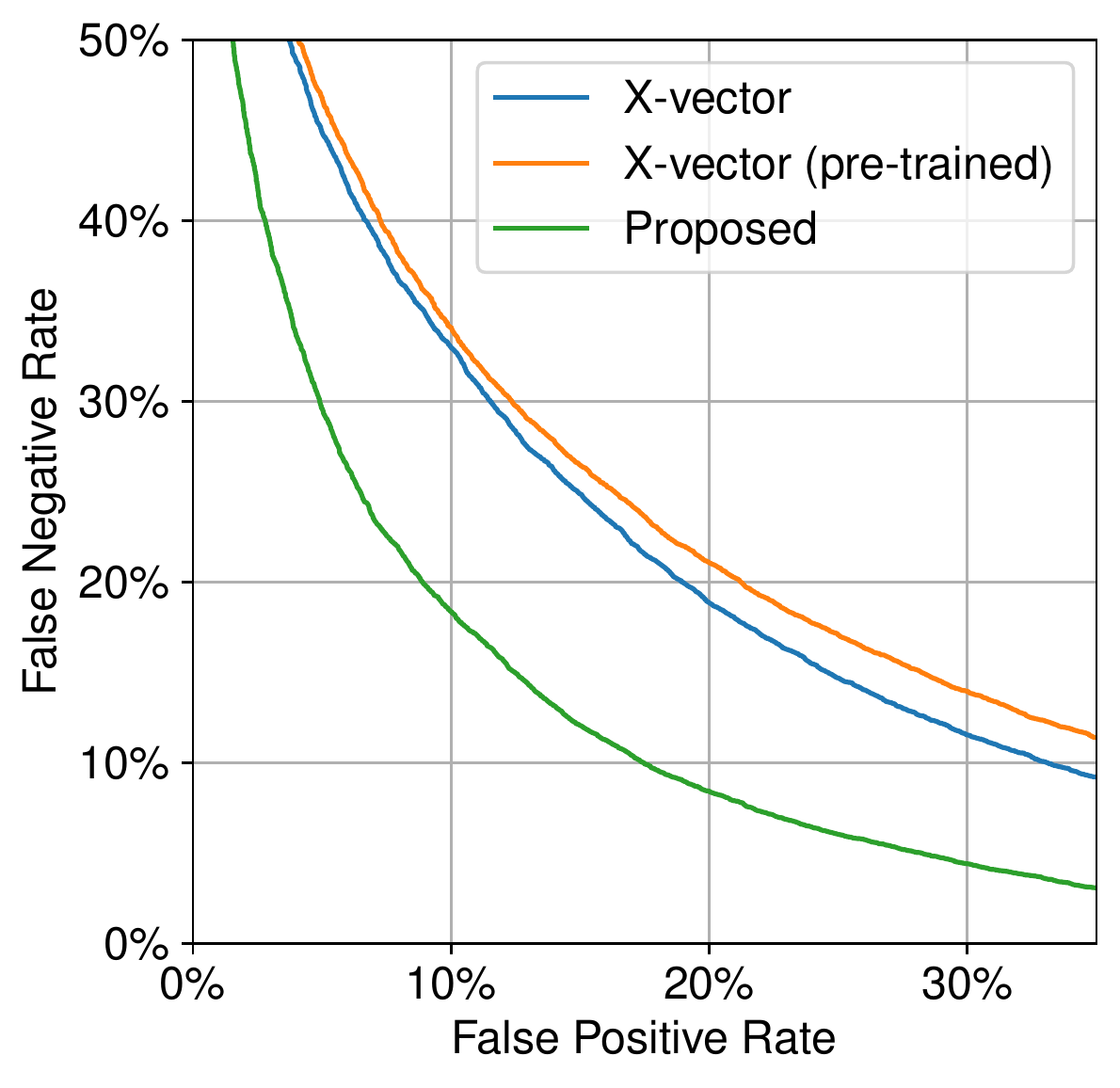}
    \label{fig:DET_I}
    \end{subfigure}
    
    \caption{\ac{DET} curves computed for the test sets $\Ra$ and $\Ia$. A performance gap is observed between the proposed system and the x-vector in the presence of interfering speakers (bottom). The curves computed for $\Na$ (omitted due to the space constraint) lies somewhere in-between $\Ra$ and $\Ia$.}
\end{figure}

The proposed model outperforms the x-vector models by a small margin on the official splits in $\Ra$. Notable gains are achieved over $\Na$ and $\Ia$ compared to the baseline. Both the proposed system and baselines demonstrate a reasonable robustness to noise, with a drop of $11.4\%$ and $32.6\%$ in their respective \acp{EER} relative to $\Ra$. In the case of $\Ia$, these figures further drop to $73.9\%$ and $114.7\%$, respectively, increasing the performance gap between the proposed and the x-vector models. This gap is visualized in the \ac{DET} curves shown in \Figures{fig:DET_C}{fig:DET_I}.

While poor x-vector performance on $\Ia$ is expected, it is surprising that the trained and pre-trained models behave similarly.
For the latter, it is conceivable that the model did not generalize for $\Ia$ since the condition was unseen in training.
This, however, is not the case for the trained x-vector; every training batch was augmented by samples containing interfering speakers.
Yet, the x-vector's capacity to accommodate interfering speakers did not improve albeit explicitly trained to fit this scenario.
This finding supports our hypothesis: learning robust speaker embeddings is challenging in a multi-speaker environment.

For the other cases, namely $\Ra$ and $\Na$, the performance similarity is expected. Both models (trained and pre-trained x-vector) reasonably generalize to noisy environments as noise augmentation is part of the training of both models.
The slight difference in the \acp{EER} can be understood due to the difference in the batch size, data preparation, and the seed.

\acresetall
\section{Conclusion}
\label{sec:conclusion}
We introduced a \acl{SV} system a multi-speaker environment using temporal feature fusion and a reference of the target speaker. The fusion of the target information with the frame-level features of the mixture allows detecting whether the target exists in the mixture or not.
We have shown that the proposed system is generally more robust against interfering signals than the x-vector baseline. The performance of the x-vector was shown to suffer in the presence of concurrent speakers, even when explicitly trained to generalize to this scenario. We have demonstrated that the proposed \acl{SV} system achieves a $30.3\%$ relative improvement over the x-vector in multi-speaker environments and $13.7\%$ on the official VoxCeleb1 splits.

\bibliographystyle{IEEEtran}
\balance
\bibliography{bibliography}

\end{document}